\pdfoutput=1
\documentclass[12pt]{article}
\usepackage{graphicx}
\usepackage{pdfsync}
\def\a{\alpha}

\def\d{\delta}
\def\e{\epsilon}

\def\f{\phi}

\def\vf{\varphi}

\def\m{\mu}
\def\n{\nu}

\begin{document}

\thispagestyle{empty}
\setcounter{page}{0}
\renewcommand{\theequation}{\thesection.\arabic{equation}}

{\hfill{ULB-TH/12-09}}

\vskip 2cm

\begin{center}
 \Large {\bf   Symmetry breaking and the Scalar boson\\ - evolving perspectives}\footnote{Invited talk presented at ``Rencontres de Moriond 2012'', March 7.}

\vskip 1cm

 \normalsize  {Fran\c cois Englert}

{\footnotesize\em Service de Physique Th\'eorique\\ Universit\'e Libre de   Bruxelles,
  Campus Plaine, C.P.225\\ and\\The International
Solvay Institutes, Campus Plaine C.P. 231\\ Boulevard du Triomphe, B-1050 Bruxelles, Belgium}\\
 {\tt
fenglert@ulb.ac.be}
\end{center}

\vskip 2.5cm

\begin{quote}
\small
\centerline{\bf Abstract}

The mechanism extending spontaneous symmetry breaking to gauge fields had a considerable impact on both theoretical and experimental elementary particle physics. It is corroborated by the discovery of the Z and the W, and by the precision electroweak tests. A detection of its Scalar boson(s) would not only constitute a direct verification of the mechanism, but knowledge of its couplings to known particles could pave the way to the world hitherto hidden beyond the Standard Model. These topics are discussed with emphasis on conceptual issues.

\end{quote}

\newpage
\tableofcontents

\baselineskip 18pt

\centerline{xxxxxxxxxxxxxxxxxxx}
\vskip 1cm

\setcounter{equation}{0}
\addtocounter{footnote}{-1}

\section{Spontaneous  breaking of a global symmetry}
\subsection{Chiral symmetry breaking}

Spontaneous symmetry breaking was introduced in relativistic quantum field theory by Nambu  in analogy with the BCS theory of superconductivity~\cite{nambu1}.   The problem studied by Nambu~\cite{nambu2} and Nambu and Jona-Lasinio~\cite{nambujl} is the spontaneous breaking of the chiral $U(1)$ symmetry of massless fermions resulting from the arbitrary relative (chiral) phase between their decoupled right and left constituent neutrinos. They then generalize to include isospin. Fermion mass cannot be generated from a chiral invariant interaction in a perturbation expansion but may arise through  a  (non-perturbative) self-consistent fermion condensate: this breaks the chiral symmetry  spontaneously. Nambu~\cite{nambu2} showed that such spontaneous symmetry breaking (SSB) is accompanied by a massless pseudoscalar. This is interpreted as the chiral limit of the (tiny on the hadron scale) pion mass. Such interpretation of the pion
constituted a breakthrough in our understanding of strong interaction physics. In the model of reference~\cite{nambujl}, it is shown that SSB also generates a massive Scalar boson, which I denote by a capital to emphasize its role in what follows.

\subsection{The simple Goldstone U(1) model}

The significance of the massless boson(s) and of the massive Scalar boson(s) occurring in  SSB is well illustrated in a simple model devised by Goldstone~\cite{goldstone}.  A complex scalar field  $\phi$ experiences a potential $V(\phi^*\phi)$.  The Lagrangian density,
\begin{equation}
\label{global} {\cal L} =\partial ^\mu\phi^*\partial_\mu\phi -V(\phi^*\phi)
\quad\hbox{with} \quad V(\phi^*\phi)= -\mu^2 \phi^*\phi  + \lambda (\phi^*\phi)^2\qquad\lambda >
0\, ,
\end{equation}  is invariant under the   $U(1)$ group $\phi\to
\displaystyle {e^{i\alpha} \phi}$.  The $U(1)$ symmetry is spontaneously broken  by the expectation
value of the
$\phi$-field acquired, at the classical level, at the minimum of the potential
$V(\phi^*\phi)$ depicted in Fig.1. 

\vskip .5cm
\begin{figure}[h]
   \centering
   \includegraphics[width=8cm]{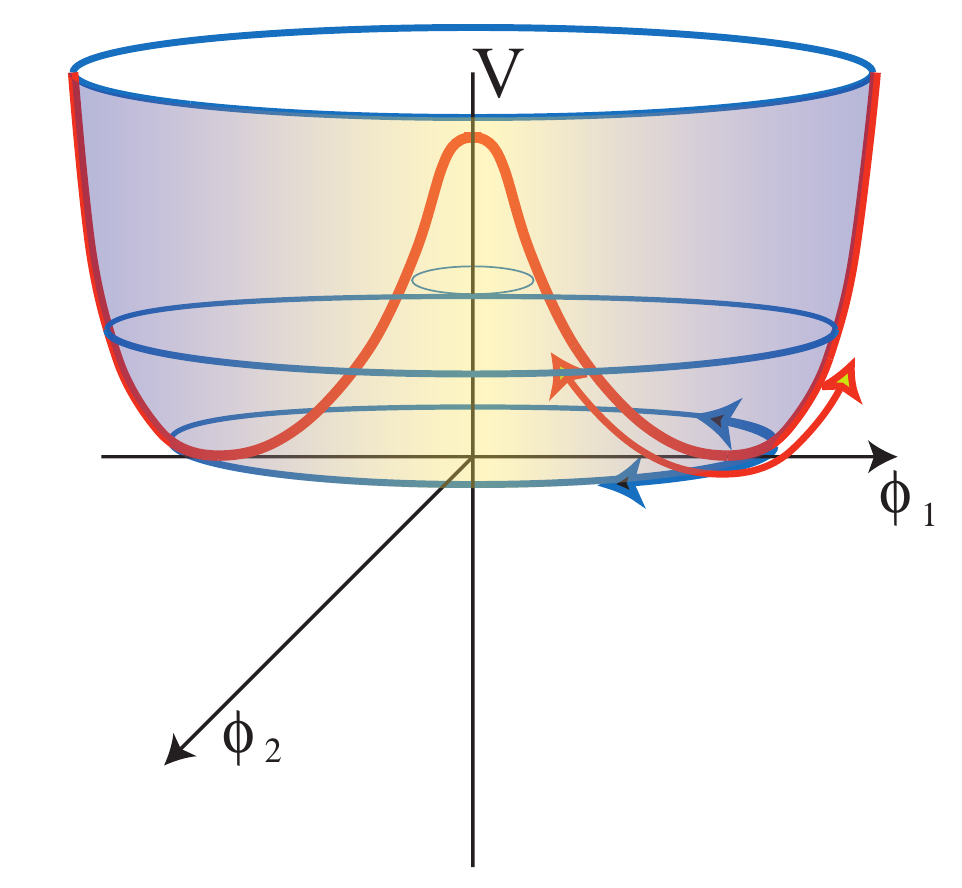}
 \caption { \footnotesize Spontaneous symmetry breaking  in the Goldstone model.}
  \end{figure}

\noindent
Writing $\phi=\langle \phi \rangle +\varphi$ and $\phi = (\phi_1 + i\phi_2)/ \sqrt2$, the $U(1)$ symmetry breaking is revealed by selecting the expectation value $\langle \phi \rangle$ to lie in some direction, say $\phi_1$, of the $(\phi_1,\phi_2)$ plane. The quadratic terms in  $\varphi_1$ and  $\varphi_2$ yield the mass squared of their respective fields, namely, using  the condition $ \langle
\phi_1\rangle^2=\mu^2/2\lambda $ at the minimum, 
\begin{equation}
\label{masses} 
m^2_{\varphi_1}= 2\mu^2\qquad m^2_{\varphi_2}= 0\,  .
\end{equation}

Thus $\varphi_2$ describes a massless  boson, $\varphi_1$  a massive one, and the ``order parameter" $\langle \f_1\rangle$ may be viewed as a condensate of $\vf_1$ bosons. Their significance is brought to light in Fig.2 and Fig.3 depicting respectively classical  $\varphi_1$ and  $\varphi_2$ waves on the background $\langle \f_1\rangle$.
\begin{figure}[h]
   \centering
   \includegraphics[width=9
  cm]{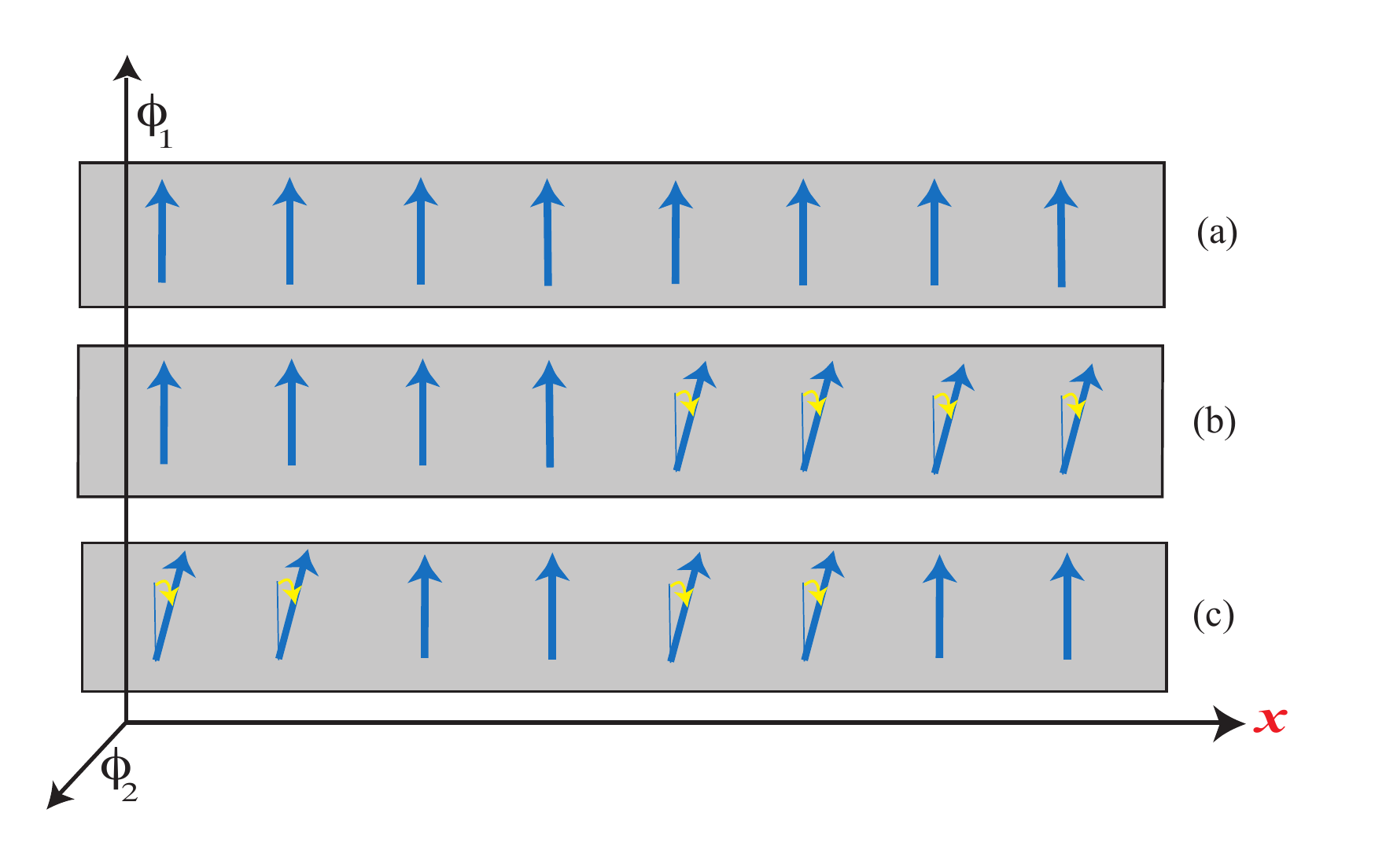}
 \caption { \footnotesize Massless Nambu-Goldstone mode $\vf_2$.}
   \label{NG}
\end{figure}

\noindent
Fig.2a represents schematically a lowest energy state (a ``vacuum") of the system: a constant non-zero value of the field $\f_1 =\langle \f_1\rangle$ pervades space-time. Fig.2b depicts the excitation resulting from the rotation of half the fields in the $(\f_1,\f_2) $ plane. This costs only an energy localized near the surface separating the rotated fields from the chosen vacuum. SSB implies indeed that rotating all the fields would cost no energy at all: one would merely trade the initial chosen vacuum for an equivalent one with the same energy. This is the characteristic {\it vacuum degeneracy} of SSB. Fig.2c mimics a wave of $\vf_2$. Comparing 2c with 2b, we see that as the wavelength of the wave increases indefinitely, its energy tends to zero, and may be viewed as generating in that limit a motion along the valley of Fig.1. Quantum excitations carried by the wave reach thus zero energy at zero momentum and the mass $m_{\vf_2}$ is zero, in agreement with Eq.(\ref{masses}). Fig.2 can easily be generalized to more complex spontaneous symmetry breaking of {\it continuous} symmetries. Massless bosons are thus a general feature of such SSB already revealed by Nambu's discovery of the massless pion resulting from spontaneous chiral symmetry breaking. They will be labeled {\it massless Nambu-Goldstone (NG) bosons}. Formal proofs corroborating the above simple analysis can be found in the literature~\cite{gsw}.

\noindent
Fig.3 depicts similarly a classical wave corresponding to a stretching of the vacuum fields. These excitations in the $\f_1$ direction describe fluctuations of the order parameter $\langle\f_1\rangle$. They are volume effects and their energy does not vanish when the wavelength becomes increasingly large. They correspond in Fig.1 to a climbing of the potential.  The quantum excitations $\vf_1$ are thus now massive, in agreement with Eq.(\ref{masses}). These considerations can be again extended to more general SSB (even to discrete ones) to account for order parameter fluctuations. Lorentz invariance imposes that such massive excitations are necessarily scalar particles. They were also already present in reference \cite{nambujl} and will be denoted in general as {\it massive Scalar bosons}. 
\begin{figure}[h]
   \centering
   \includegraphics[width=9cm]{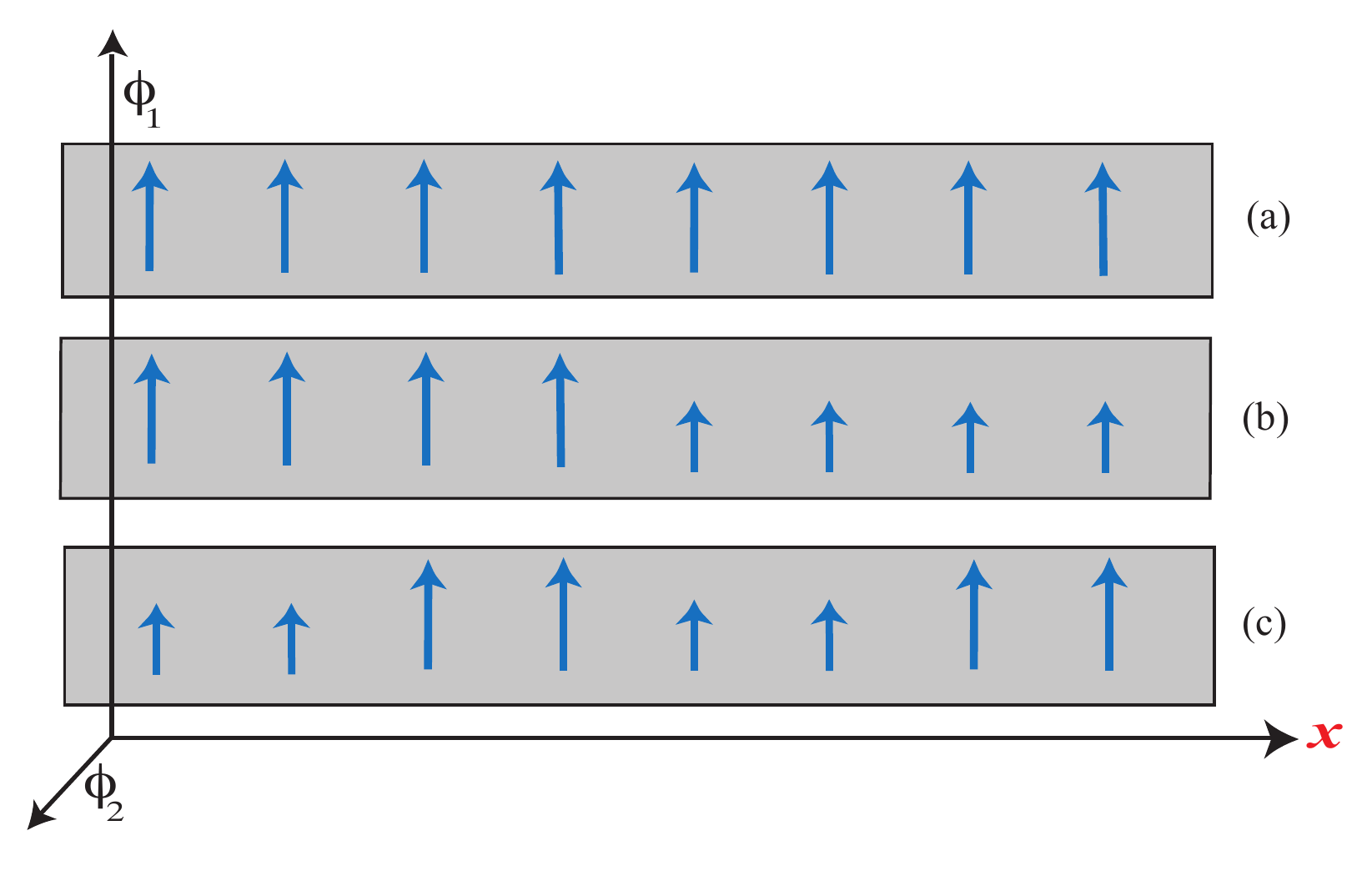}
 \caption { \footnotesize Massive Scalar mode $\vf_1$.}
  \end{figure}

The above considerations are restricted to spontaneous symmetry breaking of {\it global} continuous symmetries. Global means that the symmetry acts everywhere in space-time: for instance in the $U(1)$ Goldstone model the parameter $\a$ in $\phi\to
\displaystyle {e^{i\alpha} \phi}$ is independent of the space-time point $x$. We now discuss the extension from global to local symmetries.

\setcounter{equation}{0}

\section{The symmetry breaking mechanism for gauge fields}

\subsection{From global to local symmetry}

The global $U(1)$ symmetry in Eq.(\ref{global})  is extended  to a local one $\phi(x)\to
\displaystyle {e^{i\alpha(x)} \phi(x)}$ by introducing a vector ``gauge field"
$A_\mu(x)$ transforming under such local ``gauge transformations" as  $A_\mu(x)\to A_\mu(x) + (1/ e)
\partial_\mu \alpha(x)$. The  Lagrangian density becomes
\begin{equation}
\label{local} {\cal L} =D^\mu\phi^*D_\mu\phi -V(\phi^*\phi) -{1\over4} F_{\mu\nu} F^{\mu\nu} \, ,
\end{equation} 
where in Eq.(\ref{global}) one replaces $\partial_\m$ by the ``covariant derivative" $D_\mu\phi =
\partial_\mu\phi -ieA_\mu
\phi $ and introduces the gauge invariant field strength $F_{\mu\nu} = \partial_\mu A_\nu -\partial_\nu A_\mu $ to account for the kinetic energy of the gauge field.

Local invariance under a semi-simple Lie group $\cal G$ is realized by extending  the Lagrangian 
Eq.(\ref{local}) to incorporate  ``non-abelian"
 Yang-Mills gauge vector fields $A_\mu^a$. These transform under infinitesimal transformations of the group as  $\,\d A_\m^a(x)= \e^c (x) f_{acb}A_\m^b(x) + (1/e)\partial_\m \e^a(x)$ where $f_{acb}$ are structure constants. One gets 
\begin{eqnarray}
\label{localym} &&{\cal L}_{\cal G} =  ( D ^\mu\phi)^{*A} (D_\mu\phi)^A -V  -\displaystyle { {1\over
4}} F_{\mu\nu}^aF^{a\,\mu\nu}\, ,\\
\label{covariant} &&(D_\mu \phi)^A =\partial_\mu \phi^A -  eA_\mu^a T^{a\, AB }\phi^B\quad \quad
F_{\mu\nu}^a =\partial_\mu A_\nu^a -\partial_\nu A_\mu^a -e f^{abc}A_\mu^b A_\nu^c\,
.\qquad
\end{eqnarray}
 Here, $(D_\mu \phi)^A$ are covariant derivatives, $F_{\mu\nu}^a$ are field strengths and $\phi^A$ belongs to the representation of
$\cal G$ generated by
$T^{a\,AB }$. The potential $V$ is invariant under  $\cal G$.

The local abelian or non-abelian gauge invariance of Yang-Mills theory hinges {\em apparently}  upon
the massless character of the gauge  fields $A_\mu$, hence on the long-range character of the forces
they transmit,  as the addition of a mass term for $A_\mu$ in the Lagrangian Eq.(\ref{local}) or
(\ref{localym})  destroys gauge invariance. But  short-range forces such as the weak interaction forces  seem to be as fundamental as the electromagnetic ones. To reach a basic description of such forces one is  tempted to link this
fact  to gauge fields masses
 arising from spontaneous broken  symmetry.  However the problem of  SSB is  very different for global
and for local symmetries.

\subsection{The mechanism}
This Section is based on the field-theoretic approach of reference~\cite{eb}. In view of slips often made about the content and the dates of the 1964 papers quoted in this Section 2.2, references to these papers are detailed~\cite{all}.

\subsubsection{Breaking by Scalars}
Let us first examine the abelian case $U(1)$ as realized by the complex scalar field $\phi$  exemplified in
Eq.(\ref {local}). The interaction between the complex scalar field $\phi$ and the gauge field $A_\mu$ is
\begin{equation}
\label {interact} -ie \left(\partial_\mu \phi^* \phi - \phi^*\partial_\mu \phi\right) A^\mu + e^2 A_\mu A^\mu \phi^* \phi\, .
\end{equation}
As in the Goldstone model of Section 1.2, the SSB Yang-Mills phase is realized by a non vanishing expectation value for $\phi= (\phi_1 +i\phi_2)/\sqrt 2$, which we choose to be in the $\phi_1$-direction. Thus 
\begin{equation}
\label{vee1} 
\phi=\langle \phi \rangle +\varphi\, ,
\end{equation}
with $\phi_1 = \langle \phi_1\rangle + \vf_1$ and $\phi_2= \vf_2$, where as previously $\vf_2$ and $\vf_1$ are respectively the NG massless boson and the massive Scalar boson. 

In the covariant gauges, the free propagator of the  field
$A_\mu$ is
\begin{equation}
\label {dabelian} D_{\mu\nu}^0 ={g_{\mu\nu}-q_\mu q_\nu /q^2\over q^2} + \eta {q_\mu q_\nu/q^2
\over q^2}\, ,
\end{equation} where $\eta$ is a  gauge parameter. In what follows, we shall choose the Landau gauge defined by $\eta =0$.

The polarization tensor $\Pi_{\mu\nu}$ of the gauge field in lowest order perturbation theory around the self-consistent vacuum is given by the tadpole graphs of Fig.4,
\begin{figure}[h]
   \centering
   \includegraphics[width=8cm]{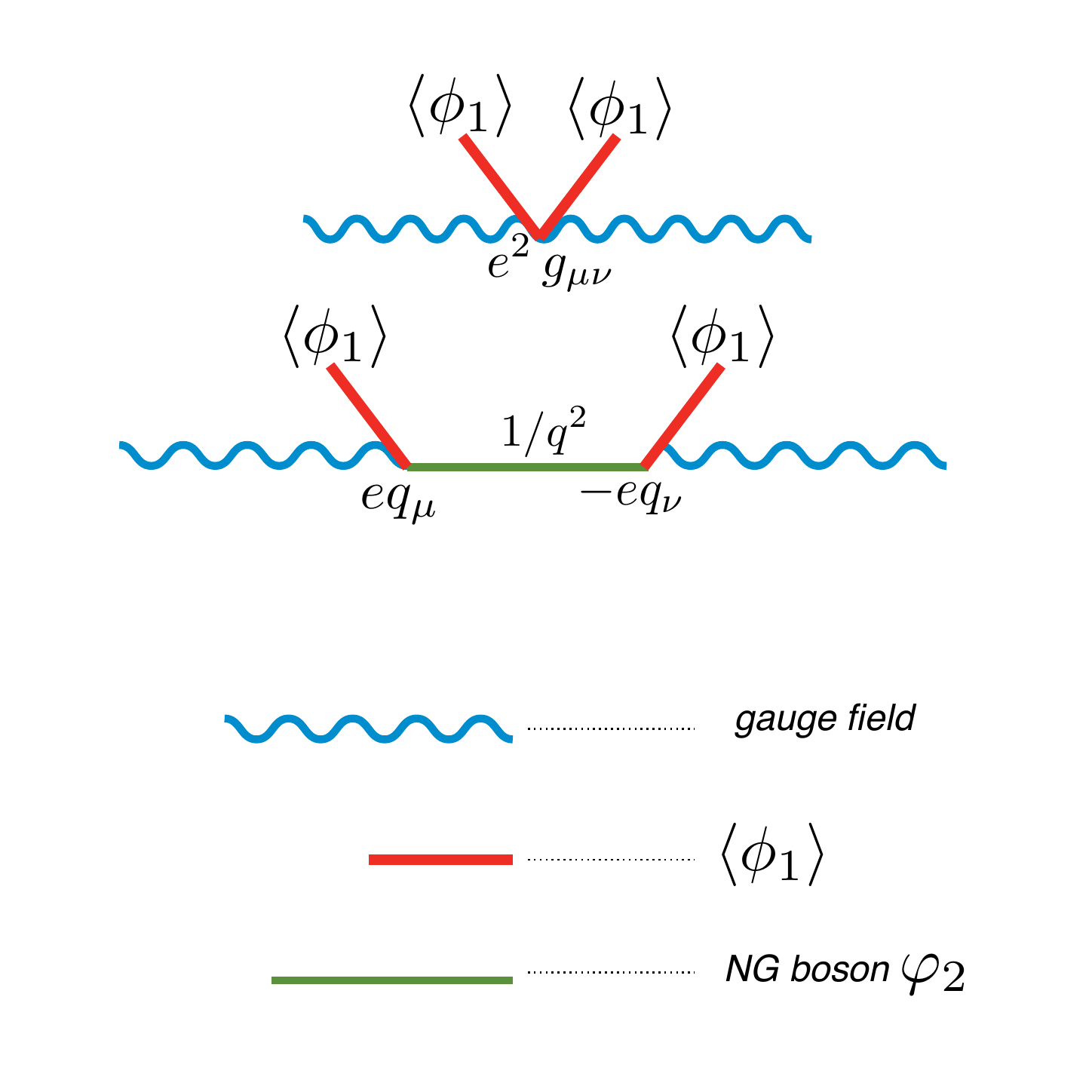}
 \caption { \footnotesize  Tadpole graphs  of SSB. Abelian gauge theory.}
   \end{figure}
 
\noindent
We see that, as a consequence of the contribution from the NG boson, the polarization tensor is transverse 
\begin{equation}
\label{transverse}
\Pi_{\mu\nu}=(g_{\mu\nu} q^2-q_\mu q_\nu) \Pi(q^2)\, ,
\end{equation}
and yields a singular
 polarization scalar $\Pi(q^2)$ at $q^2=0$,
 \begin{equation}
\label{pamass}
\Pi(q^2)={e^2\langle \phi_1\rangle^2\over q^2}\, .
\end{equation}
From Eqs.(\ref{dabelian}), (\ref{transverse}) and (\ref{pamass}),   the dressed gauge  field propagator becomes
\begin{equation}
\label{damass} D_{\mu\nu} ={g_{\mu\nu}-q_\mu q_\nu/q^2
\over q^2- M_V^2} \, , 
\end{equation} 
which shows that the $A_\mu$-field gets a mass $M_V$,
\begin{equation}
\label{massa}
M_V^2=e^2 \langle
\phi_1\rangle^2\, .
\end{equation}
The transversality of the polarization tensor Eq.(\ref{transverse}) results from the contribution of the NG boson and agrees with a Ward identity which guarantees that gauge invariance is preserved. Thus the mass of the gauge field $A_\m$ acquired  through the absorption of the NG boson is gauge invariant.

The generalization of these results to the non-abelian case described by
the action Eq.(\ref{localym}) is straightforward. Writing the generators in terms of the real components of the fields, one gets the mass matrix
\begin{equation}
\label{massna}
(M_V^2)^{ab} = - e^2\langle \phi^{B}\rangle T^{a\,BC}T^{b\,CA}\langle\phi^A\rangle\, , 
\end{equation} 
and the dressed gauge boson propagators have the same form as Eq.(\ref{damass}) in terms of the diagonalized mass matrix. 
As in the abelian case, the would-be NG bosons are absorbed by the gauge fields and  generate gauge invariant masses  in $\cal
G/H$. Long-range forces only survive in the subgroup
$\cal H$ of
$\cal G$ which leaves invariant the non-vanishing expectation values
$\langle\phi^A\rangle$.

The introduction of gauge fields and hence local symmetries resulted in the absorption of the NG boson in the gauge field propagator and in the generation of  gauge field mass. These results are encoded in Eqs.(\ref{transverse}), (\ref{massa}) and (\ref{massna}). Such consequences of local symmetry seems at odd with the appearance of massless NG bosons in global symmetries and calls for an elucidation of the concepts involved in extending the symmetry from global to local. This will now be done by unraveling the significance of the results of Section 2.2.1 for the NG boson and for the Scalar boson. To avoid notational complications, I shall mostly consider the $U(1)$ extension from the global Goldstone model to its local counterpart, although the discussion in the Sections below apply in general to the non-abelian case as well .

\subsubsection{The fate of the massless NG boson}

The diagrams of Fig.4 show that the NG boson is absorbed in the gauge field propagator. This yields the required longitudinal  polarization of the massive gauge field encoded in the numerator of Eq.(\ref{damass}) on the mass shell $q^2 = M_V^2$.  The massless  NG boson actually disappears entirely from the physical spectrum. This is an immediate consequence of  gauge invariance. Consider indeed Fig.2. As explained in Section~1.2, the massless NG mode originates in global SSB from the vacuum degeneracy: the energy of the excitations depicted in Fig.2b and 2c tend to zero in the limit of infinite wavelength because they generate  in that limit a vacuum equivalent  to the original one under a symmetry operation. But local symmetry means that the configurations of Fig.2b and 2c carry no energy at all ! They are thus simply redundant description of the {\it same} gauge invariant vacuum, a redundancy not unexpected when fields are described by potentials $A_\m$. {\it Therefore there is no vacuum degeneracy, no spontaneous symmetry breaking and thus no massless NG boson !} \footnote{A more detailed description of the distinctive features of global and local SSB can be found in reference~\cite{e1}.  Formal proofs for the absence of massless NG bosons were given by Higgs~\cite{higgs1}, and then by Guralnik, Hagen and Kibble~\cite{GHK}. These proofs do not make use explicitly of the unicity of the gauge invariant vacuum.}

An apparent symmetry breaking, akin to  the Goldstone model $U(1)$ SSB, appears when one {\it chooses} a fixed orientation of the average Scalar field $\langle \f \rangle$, e.g. $\langle \f \rangle=\langle \f_1 \rangle$. But this description is only a {\it convenient} gauge choice. It allows for the conventional assignment of group quantum numbers (such as isospin) to particles in perturbation theory. I shall therefore qualify also as SSB the mechanism generating mass for gauge fields but one should keep in mind that the symmetry is not intrinsically broken, a fact that renders the disappearance of massless NG bosons obvious. Their degrees of freedom are recovered in the longitudinal polarization of the massive gauge fields\footnote{A non relativistic precursor of this effect was found by Anderson~\cite{anderson} in condensed matter physics. Namely in superconductivity the massless mode of the broken $U(1)$ symmetry disappears by being absorbed by electron density oscillations, namely by the longitudinal ``massive" plasma mode.}.

\subsubsection{The fate of the massive Scalar boson}

A glance at Fig.3 shows that the stretching of (classical) Scalar fields are independent of local rotations of the $\f$-field in the $(\f_1,\f_2)$ plane. This translates the fact that the modulus of the $\f$-field is gauge invariant. Hence the Scalar bosons survives the gauging and their classical analysis is identical to the one given for the Goldstone model in Section~1.2.
\begin{figure}[h]
   \centering
   \includegraphics[width=12cm]{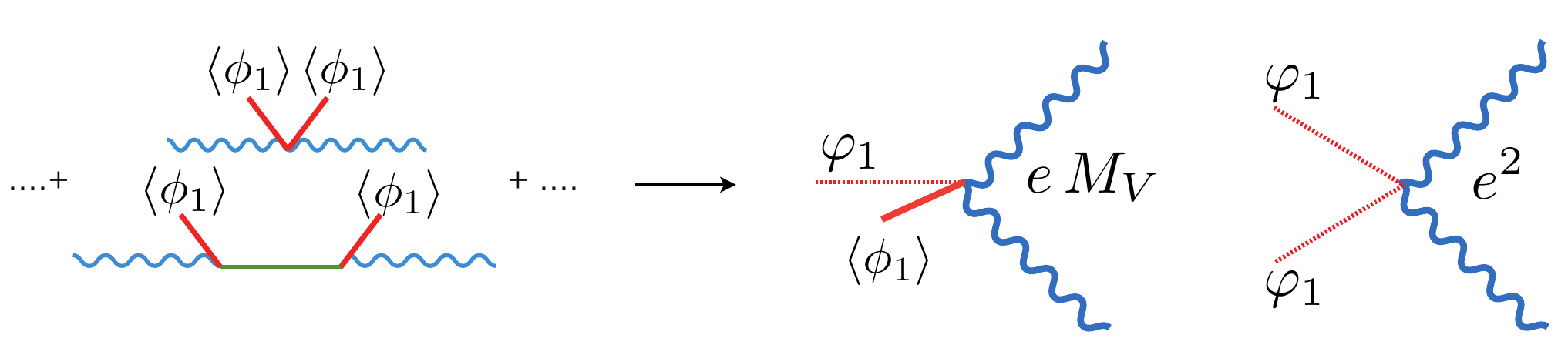}
 \caption {\footnotesize Coupling of the Scalar boson $\vf_1$ to massive gauge bosons.}
  \end{figure}

The coupling of the Scalar boson to the massive gauge bosons follows from the graphs in Fig.4. Using Eq.(\ref{vee1}) one gets the two tree-level vertices of Fig.5 where the heavy wiggly lines on the right hand side represent (tree-level) dressed massive gauge propagators. The vertex couplings follow from Eq.(\ref{massa}).

\subsubsection {Dynamical symmetry breaking}

The symmetry breaking giving mass to gauge vector bosons may also arise from a fermion condensate. If  a spontaneously broken global symmetry is extended to a local one by introducing gauge fields, the massless NG bosons disappear as previously from the  physical spectrum and their absorption by gauge fields renders these massive. 

 \subsubsection{The renormalization issue}
 
 The interest in the symmetry breaking mechanism stems from the fact that it provides, as does quantum electrodynamics, a taming of quantum fluctuations. This allows the computation of the quantum effects necessary to cope with precision experiments. In other words, the theory is ``renormalizable", in contradistinction to the theory of genuine non-abelian massive vector fields. 
 
The massive vector propagator Eq.(\ref{damass}), which is also valid in the non-abelian case by diagonalizing the mass matrix Eq.(\ref{massna}), differs from a conventional free massive vector propagator. The numerator of the former is transverse for all momenta while the  numerator of the latter, $g_{\m\n}-q_\m q_\n/M_V^2$, is only transverse on the mass shell $q^2=M_V^2$. The soft behavior at large $q^2$ of the propagator Eq.(\ref{damass})  and the gauge invariance condition Eq.(\ref{transverse}) are reminiscent of quantum electrodynamics.
This suggested that the SSB mechanism renders charge vector meson theories 
renormalizable~\cite{eb2}.

However there is a catch. The pole at $q^2=0$ in Eq.(\ref{damass})  has a negative residue and therefore is potentially  violating unitarity. A glimpse into the solution of the problem appears from comparing our approach to the one of Higgs~\cite{higgs2}. Higgs obtained most of our results from the classical equations of motion. In addition, he
showed how to eliminate all contributions of the massless NG boson in that limit by the following field transformation
\begin{equation}
\label{hvector}
  A_\mu - {1\over e\langle
\phi_1\rangle}\partial_\mu 
\phi_2 =B_\mu \, ,
\end{equation}
where $ B_\mu$ satisfies the conventional classical equations of motion of a massive vector field. In terms of propagators Eq.(\ref{hvector}) becomes the {\it identity}
\begin{equation}
\label{damass2} {g_{\mu\nu}-q_\mu q_\nu/q^2\over q^2- M_V^2}  -{1\over M_V^2} {q_\mu q_\nu
\over q^2}  = {g_{\mu\nu}-q_\mu q_\nu/M_V^2
\over q^2- M_V^2}  \, .
\end{equation}
The term in the right hand side of Eq.(\ref{damass2}) is indeed the conventional 
massive vector propagator of Higgs $B_\m$-field which displays no unwanted pole at $q^2=0$. It constitutes a ``unitary gauge'' propagator.  It does not share the soft high $q^2$ behavior of the ``renormalizable gauge" propagator Eq.(\ref{damass}). Gauge invariance should allow the use of either propagator, and the theory is thus expected to be both renormalizable and unitary. How can this happen?

The answer  lies in the second term of Eq.(\ref{damass2}). Let us couple Eq.(\ref{damass}) to an external (non-conserved) current associated to the SSB gauge symmetry. The second term in Eq.(\ref{damass2}) describes the   coupling of the Goldstone boson to its divergence.  Note that the pole contribution of the Goldstone is cancelled on-shell by the unphysical $q^2$ pole of the propagator Eq.(\ref{damass}), leaving only off-shell contributions in agreement with the fact that the massless Goldstone boson has to disappear from the physical spectrum. Thus the identity Eq.(\ref{damass2}) indicates that  the off-shell contributions of the Goldstone are needed to restore unitarity in the renormalizable gauge. As an example, one easily verifies at the tree level that, taking into account the Goldstone contribution,  the identity Eq.(\ref{damass2}) ensures the equivalence of the renormalizable and the unitary gauges in the electroweak theory discussed below.

Although these arguments suggest that the mechanism can be consistent, it is a highly non trivial affair to show that the fully interacting theory  is renormalizable and unitary.  This was  proven by 't Hooft and Veltman~\cite{renorm}, who thereby established the quantum consistency  of the SSB mechanism\footnote{See also reference~\cite{lzj}.}. 

\setcounter{equation}{0}
\section{The electroweak theory}

In the electroweak theory for weak and electromagnetic interactions~\cite{gws}, the gauge group is taken to be the chiral group
$SU(2)\times U(1)$ with $SU(2)$ acting on left-handed fermions only. The corresponding generators and coupling constants are
$gA_\mu^a\,T^a$ and $g^\prime B_\mu \,Y/2 $. 
 The Scalar field $\phi$ is a doublet of $SU(2)$ and its
$U(1)$ charge is $Y = 1$.  Breaking follows from a Goldstone type potential. It is  characterized by $
\langle\phi\rangle={1/\sqrt2}\  \{ 0, v\} $ and $Q=T^3 + Y/2$ generates the  unbroken
subgroup.  $Q$ is identified with the electromagnetic charge
operator. The only residual massless gauge boson is identified with the photon
and the electric charge $e$ is usually expressed in terms of the mixing angle
$\theta$ as $g=e/\sin\theta  , g^\prime=e/\cos\theta $. The expectation value $\langle\phi\rangle$ generates the masses of all known elementary fermions through Yukawa couplings.

Using Eqs.(\ref{massa}) and (\ref{massna}) one gets the mass matrix
\begin{center}
$\vert\mu^2\vert$=$\displaystyle{{v^2\over 4}} ~\begin{array}{|cccc|} g^2&0&0&0\\ 0&g^2&0&0\\
0&0&g^{\prime 2}& -g g^\prime\\ 0&0&-g g^\prime & g^2
\end{array}$
\end{center} whose diagonalization yields the eigenvalues
\begin{equation} M^2_{W^+}={v^2\over 4}g^2\qquad M^2_{W^-}={v^2\over
4}g^2\qquad M^2_Z={v^2\over 4}~(g^{\prime 2}+ g^2)\qquad M^2_A= 0 \, . 
\end{equation}

The discovery of the $Z$ and $W$ bosons in 1983 and the precision experiments testing the quantum consistency of the theory establish the validity of the mechanism. The Scalar boson itself is presently search for and would provide a direct proof of it and would also characterize its precise realization. This will be further discussed in the following Section. The couplings of the Scalar to the massive $W$ and $Z$ bosons follow from Fig.5 and are depicted in Fig.6a. Its coupling to elementary fermions similarly follows from the Yukawa couplings and are shown in Fig.6b. The coupling to the massless photons occur at the loop level as indicated in Fig.7.

\vskip 1cm
\begin{figure}[h]
   \centering
   \includegraphics[width=12cm]{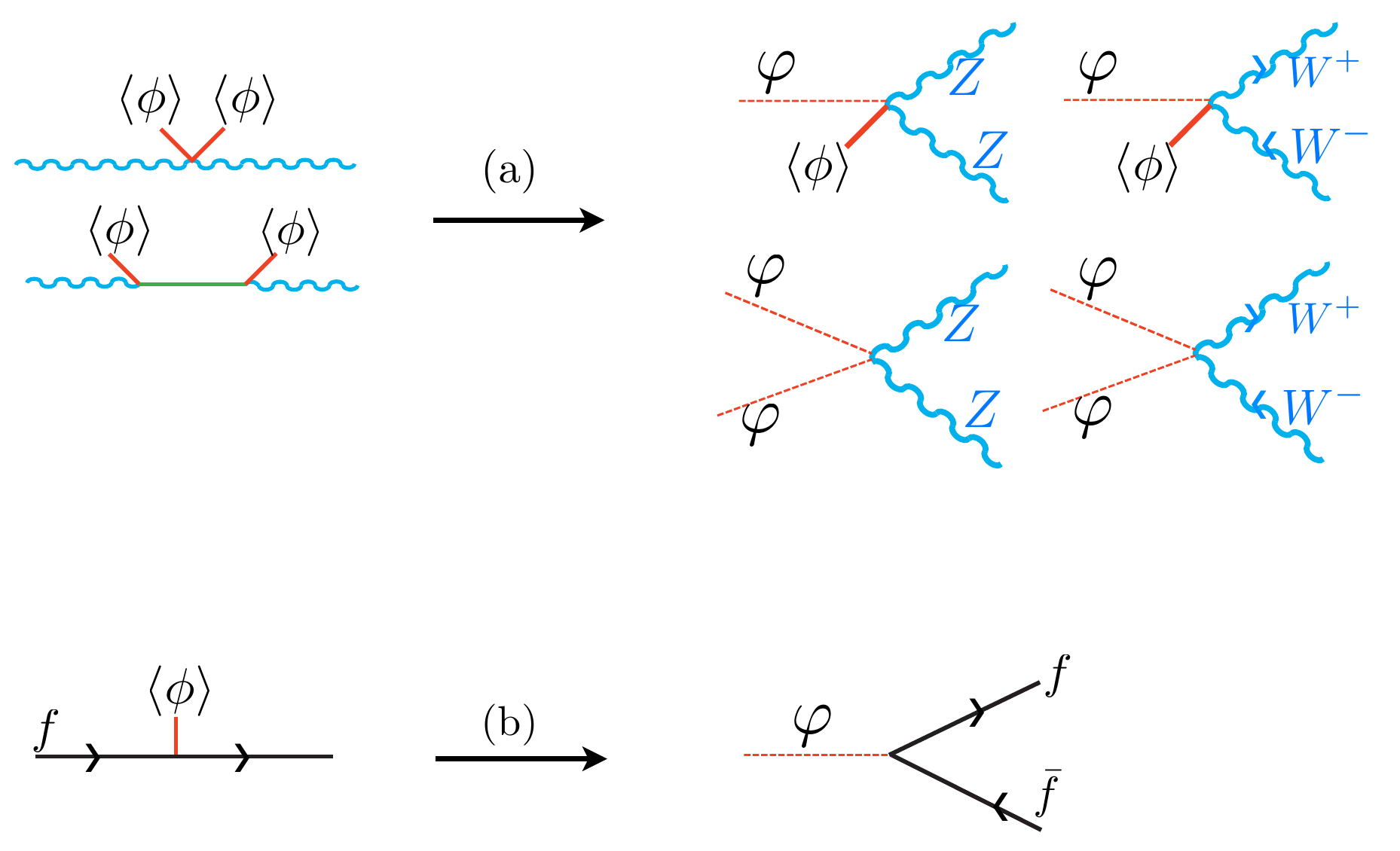}
 \caption {\footnotesize Coupling of the Scalar boson $\vf$ to massive gauge bosons and to elementary fermions.}
 \end{figure}

\begin{figure}[h]
   \centering
   \includegraphics[width=12 cm]{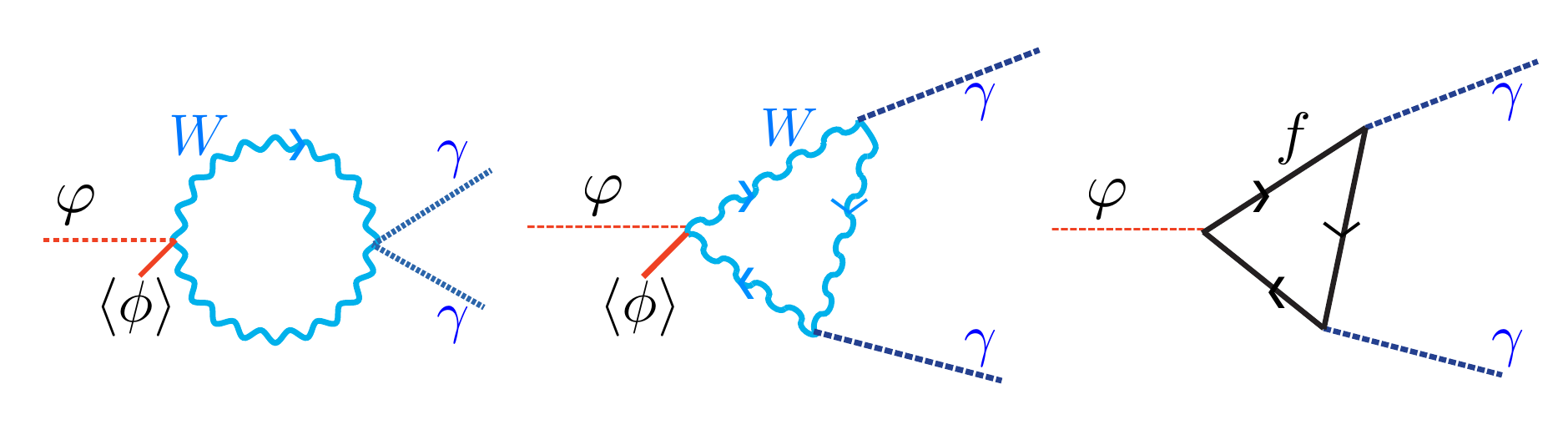}
 \caption {\footnotesize  Coupling of the Scalar boson $\vf$ to photons.}
 \end{figure}

\section {Perspectives}

Hopefully a Scalar boson will be found. Will it be identifiable to the electroweak Scalar of the Standard model, at least up to energies presently available at the LHC ? and what would be hidden beyond the Standard Model at possible higher energies? Although the answer must obviously be deferred to  experiment, one may try, assuming that some Scalar will be found,  some educated guess based on the belief that the organization of nature in scientific terms should be based on a minimal number of disjoint hypothesis. The usefulness of this exercise is that properties of known particles may perhaps pave the way to  the unknown.

To try to make sense of this type of Occam's razor approach, I will select and compare some perspectives. I consider two perspectives which I find appealing. The first one I label {\it historical} :  it takes at its starting point the SSB mechanism. The second one evolves out of the requirement that elementary fermions be massive; I will  label it {\it logical}. In both perspectives, now that the chiral character of weak interactions is well established for massive fermions,  the backbone of the electroweak theory is taken to be  its group-theoretical content: the chiral group $SU(2) \times U(1)$.

\medskip
\noindent
 - a) A ``historical" perspective

The Fermi theory of weak   interactions, formulated in terms of a four Fermi point-like current-current interaction, was well-defined in lowest order perturbation theory and successfully confronted many experimental data. However, it is clearly inconsistent in higher orders because of uncontrollable divergent quantum fluctuations. In order words, in contradistinction to quantum electrodynamics,  the Fermi theory is not renormalizable. This
difficulty could not be solved by smoothing the point-like interaction by a massive, and therefore short-range, charged vector particle exchange (the  $W^+$ and $W^-$ bosons): theories with  massive  charged vector bosons are  not renormalizable either. It is the electroweak theory, based on the renormalizable SSB mechanism applied to the group $SU(2) \times U(1)$, that provided a suitable tool for testing experiment. Its Scalar boson has yet to be confirmed.

The Scalar boson was introduced there through a field experiencing a Goldstone-type potential. In absence of a yet definite experimental answer, should this be expected to be an elementary particle (at least to testable scales) or is it a phenomenological description of a composite object?

The SSB mechanism described in Section 2 could be realized by an elementary Scalar condensate (Section 2.2.1) or dynamically (Section 2.2.4) in which case the Scalar boson would, at best, be a bound state. The absence of known elementary scalar particle, and the fact that a neutral scalar condensate could be, as is often the case in condensed matter physics,  only a phenomenological description of a more complex dynamics, may suggest a dynamical realization of the mechanism. This is comforted by the fact that simple dynamical models, such as Technicolor, can be constructed. Technicolor generate gauge vector boson masses, but its extension to produce elementary fermion masses is more problematic.  Giving mass to the fermions dynamically, which is natural in this perspective, might require additional groups which have then to confront many experimental constraints. As a rule, full dynamical symmetry breaking is very laborious and the corresponding phenomenological Scalar(s) may  have higher masses\footnote{For a review on dynamical electroweak symmetry breaking see reference~\cite{hs}.}. 

{\it The ``historical" perspective suggests that at low energies, the Scalar decays of Fig.6a (but not 6b) and Fig.7 could perhaps be presently detectable except  for the last diagram involving fermion loops. Of course similar ``fermiophobic'' constraints would apply to the creation process. At higher energies one would expect the emergence of a complex spectrum beyond the standard model one.}

\medskip
\noindent
 - b) A ``logical" perspective
 
Generating elementary fermions  mass at the outset and hence breaking the chiral symmetry of $SU(2) \times U(1)$  is a simple problem if one introduces an elementary Scalar with Yukawa couplings. Actually, this requires only global symmetry breaking and NG bosons are produced. To eliminate them one has to extend the symmetry to a local one and thus the gauge fields  of the group must be introduced. It is natural to require that the elementary Scalar giving mass to the fermions acts then on the gauge fields. Of course one just recovers  the electroweak  theory and all decays of Fig.6 and Fig.7 should occur. The simplicity of the ``logical'' perspective, as compared to the ``historical'' one, is marred by the introduction of an elementary Scalar disconnected from the fermionic content of the theory and submitted to the ad hoc Goldstone-type potential. These drawbacks could largely be avoided if some hidden supersymmetry, broken at larger energy scales, would be present. It would ensure, independent of the usual rather weak argument of ``naturalness'', that elementary scalars do appear and are accompanied by fermions at different masses.

{\it The ``logical" perspective suggests that at low energies, the electroweak theory provides the correct description. At higher energies one would expect  some kind of hidden supersymmetry. In the ``historical'' perspective there appear to be no particular reason for postulating supersymmetry.}

If there is some sense in selecting these two perspectives, it would mean that if the Scalar is indeed discovered but is ``fermiophobic", one would expect at higher energy scales no supersymmetry but instead a very complex structure with many new particles. If however the Scalar has the properties predicted by the electroweak theory, supersymmetry broken at high energy would be expected but the complexity of group and dynamical structures at high energy might be tamed. Clearly, these considerations stem from some kind of aesthetic prejudices, whatever that means, and only experiment will tell. They just convey a touch of personal feelings about the problem: I became increasingly impressed by the coherence out of the very few hypothesis embedded in the ``logical'' perspective  and I took this occasion to stress it.

\section*{Acknowledgments}
This work was supported in part by IISN-Belgium and by 
the Belgian Federal Science Policy Office through the 
Interuniversity Attraction Pole ``Fundamental Interactions''.

\noindent
I am very grateful to Jean-Marie Fr\`ere for illuminating discussions covering all parts of this presentation.

\end{document}